\documentclass[12pt,a4paper]{article}
\usepackage{amssymb,amsmath,amsthm}
\usepackage{natbib}
\theoremstyle{plain}

\theoremstyle{definition}

\numberwithin{equation}{section}
\usepackage{fullpage}

\usepackage{cite}

\usepackage{graphicx}

\begin{document}

\title{Heterogeneous Beliefs with Finite-Lived Agents}

\author{ A.A. Brown\thanks{Wilberforce Road, Cambridge CB3 0WB, UK (phone = +44 1223 337969 , email = A.A.Brown@statslab.cam.ac.uk)}  \\ \small Statistical Laboratory, \\  \small University of Cambridge \and L.C.G. Rogers\thanks{Wilberforce Road, Cambridge CB3 0WB, UK (phone = +44 1223 766806, email = L.C.G.Rogers@statslab.cam.ac.uk)} \\ \small Statistical Laboratory, \\ \small University of Cambridge }

\date{First Version: January 2008  \\ This Version: July 2009 }

\maketitle

\graphicspath{{DBOUCase/}}

\abstract{This paper will examine a model with many agents, each of whom has a different belief about the dynamics of a risky asset.  The agents are Bayesian and so learn about the asset over time.  All agents are assumed to have a finite (but random) lifetime. When an agent dies, he passes his wealth (but not his knowledge) onto his heir.   As a result, the agents never become sure of the dynamics of the risky asset.  We derive expressions for the stock price and riskless rate.  We then use numerical examples to exhibit their behaviour.}

\section{Introduction}

This paper will look at a model of agents with heterogeneous beliefs.  We assume that there is a single risky asset that produces a dividend process.  Agents are unsure of the dynamics of the dividend process.  Specifically, they do not know one of the parameters that governs its dynamics.  Agents therefore form beliefs about this parameter and update these over time.  To avoid agents eventually determining the true value of the parameter, we assume that agents are finite lived.

The paper will build on previous work of \citet{BrownRogersDiverseBeliefs}.  That paper explained the general theory of how to incorporate heterogeneous beliefs into a dynamic equilibrium model.  However, in the case in which the agents were Bayesian, it was seen that the agents would eventually determine the true drift of the dividend process.  The purpose of this paper is therefore to investigate a model in which there is a non-trivial steady state.  This is done through the assumption that the different agents are in fact dynasties.  Each member of the dynasty has a finite but random lifetime and when that member dies, he will pass on his wealth, but not his knowledge, to his heir.  The paper will explain how to construct and solve this model and will lead to a stationary distribution for the stock price.  

As in \citet{BrownRogersDiverseBeliefs}, we assume that there is a single risky asset which pays a dividend continuously in time.  In addition there is a riskless asset in zero net supply.   The dividend process of the stock is now assumed to be a quadratic function of an Ornstein-Uhlenbeck (OU) process.  All the agents know all the parameters of the OU process except the mean to which it reverts.  All the agents observe the OU process as it evolves and so as time progresses they update their beliefs about the unknown parameter.  However, since they are finite lived, they will never find its true value.

The model described is quite simple, yet already there is enough to make the asset pricing non-trivial.  Just as in \citet{BrownRogersDiverseBeliefs}, the agents maximise their expected utilities subject to their budget constraints and we use these optimisation problems to derive a state price density.  Using this state price density we can then price the risky asset as the net present value of future dividends.  Comparative statics allow us to see how the stock price depends on the parameters of our model.  We also produce a volatility surface for the stock, which behaves very reasonably.

The structure of the paper is as follows.  We give a brief literature review below.  Section \ref{DBOUCaseModel} introduces the model and solves the equilibrium to determine a state price density.  Section \ref{DBOUCaseAssetPrices} then uses this state price density to calculate the prices of the stock and bond; these calculations are non-trivial.  Section \ref{DBOUCaseNumRes} looks at comparative statics of the model and section \ref{DBOUCaseConclusions} concludes.

\subsection{Literature Review}

There is a large literature on heterogeneous beliefs, which has been discussed in detail in \citet{BrownRogersDiverseBeliefs}.  Work includes \citet{Kurz}, \citet{Kurz94}, \citet{Kurz97}, \citet{KurzMotolese}, \citet{Kurz08}, \citet{KJM}, \citet{Fan}, \citet{HarrisonKreps}, \citet{Morris}, \citet{WuGuo03}, \citet{WuGuo04}, \citet{HarrisRaviv}, \citet{KandelPearson}, \citet{BuraschiJiltsov}, \citet{JouiniNapp}.  Closer to the work presented here are the papers that assume that there is a parameter of the economy that is unknown to the agents.  We briefly review such models here.

\citet{Basak} considers a two-agent model in which each agent receives an endowment process.  There is also an extraneous process that agents believe may effect the economy.  The endowment process and all its parameters are observed.  The extraneous process is observed, but the parameters of the stochastic differential equation (SDE) that drives it are not known to the agents.  They form beliefs about the drift term in this SDE and update their beliefs in a Bayesian manner.  The paper analyses this problem and derives quantities such as the consumption, the state price density and riskless rate.  Basak also explains how to generalise the model to multiple agents and multiple extraneous processes. 

\citet{Basak2005} also considers a model with two agents, who each receive an endowment process. The aggregate endowment process is observed by the agents.  They also observe its volatility, but not its drift; they use filtering to determine this drift.  There is assumed to be a bond and risky security, both in zero net supply.  Again, agents do not know the drift of the stock price.  Agents maximise the expected utility of consumption.  He then solves for the equilibrium and uses it to derive interest rates and perceived market risk of the agents.  He also gives a number of generalisations to the model.  For example, he considers the case in which there is a process which does not directly affect the asset prices.  However, each agent thinks that this process does affect the dynamics of the asset prices and so this changes the equilibrium.  He also looks at the case of multiple agents and again derives the riskless rate and perceived market prices of risk.  The final part of his paper looks at further extensions to his model; for example, he explores a monetary model in which there is a money supply that is stochastic and agents disagree on its drift.  

\citet{GallmeyerHollifield} have considered the effects of adding a short-sale constraint to a model with heterogeneous beliefs.  They consider a model with two agents.  These agents are unsure about the drift of the output process of the economy.  They start with initial beliefs about the drift and use filtering to update these.  The agent who is initially more pessimistic is assumed to have logarithmic utility and a short sale constraint.  The optimistic agent is assumed to have general CRRA utility and does not have a short sale constraint.  The authors examine this model and derive expressions for the state price densities, stock price and consumption. In particular, they examine the effects of the imposition of the short sale constraint on the stock price.

The paper of \citet{Zapatero}  considers a model in which there is an aggregate endowment process that obeys an SDE driven by two independent Brownian motions.  The constant drift of the process is unknown to the agents.  There are 2 groups of agents and they each have a different Gaussian prior for this drift.  Zapatero also considers the case in which as well as observing the endowment process, the agents also see a signal, which again is driven by the two Brownian motions, but has unknown drift. Again, agents have prior beliefs about this drift, which they update.  He derives an equilibrium and shows that volatility of the interest rate is higher in an economy with the additional information source.  

\citet{Li}  considers a model with 2 groups of agents.  There is a dividend process which obeys some SDE, but the drift of this SDE is unknown.  The drift can satisfy one of two different SDEs.  Each group of agents attaches a different probability to the drift obeying the two different SDEs.  They update this probability as they observe more data.  Agents are assumed to have log utility and Li derives the stock price, wealth and consumption of agents in this model.  He also analyses the volatility of the stock price.

Turning to the Bayesian learning side of our story, we remark that there is an extensive literature on Bayesian learning in finance and economics in which agents update their beliefs as they observe data.  Work includes \citet{HautschHess2004}, \citet{KandelPearson}, \citet{SchinkelTuinstraVermeulen}, \citet{KalaiLehrer}  each of whom uses this Bayesian learning in quite different setups.  For example, \citet{SchinkelTuinstraVermeulen} apply Bayesian learning to $n$ competitive firms who set prices but do not know the demand function.  They observe demand at each step and use this to update their posterior belief for the state of the world, which then impacts their perceived demand function. The authors show that prices converge.  \citet{KalaiLehrer} applies Bayesian learning to an $n$-person game in which agents do not know the payoff matrices of their competitors.  They show that the equilibrium will approach the Nash equilibrium of the system.  \citet{HautschHess2004} apply Bayesian learning to explain why more precise data has a larger impact on market prices.  They test this by looking at the behaviour of T-bond futures when unemployment data is announced.   

Closer to our work, \citet{GuidolinTimmermann} look at a discrete time model in which the dividend process can have one of two different growth rates over each time period and the probability of each growth rate is unknown to the agents.  The agents are learning, so update their estimate for the unknown probability at each time step.  In order to avoid the problem of agents discovering the true probability, they also consider agents who only look at a rolling window of data.

\section{The Model}\label{DBOUCaseModel}

The setup of our model is similar to \citet{BrownRogersDiverseBeliefs}.  There is a single productive asset, which we refer to as the stock, which pays dividends continuously in time. The dividend at time $t$ is $\delta_{t}$.   The dividend process is assumed to be a quadratic function of a stationary Ornstein Uhlenbeck (OU) process.  

Since we are interested in obtaining a stationary distribution for the stock price, the construction of the probability space requires slightly more care than in \citet{BrownRogersDiverseBeliefs}. Let $\Omega$ denote the sample space. We set $\Omega=C(\mathbb{R}, \mathbb{R})$, the space of continuous functions from $\mathbb{R}$ to $\mathbb{R}$.  Let $X_{t}(\omega) \equiv \omega(t)$ denote the canonical process. Furthermore, let $\mathcal{F}_{t}=\sigma({X_{s}:-\infty \le s \le t})$.  

As before, the reference measure is denoted by  $\mathbb{P}_{0}$.  We assume that under this measure $X$ is a stationary OU\footnote{An Ornstein Uhlenbeck process which reverts to mean $a'$ with reversion rate $\lambda$ satisfies the SDE $dX_{t} = d\tilde{W}_{t} - \lambda (a'-X_{t}) dt $ where $\tilde{W}$ is a standard Brownian motion under the reference measure.}  process which reverts to mean zero and has reversion rate $\lambda$. 

Next, we define:
\begin{align}
W_{t} = X_{t} - X_{0} + \int_{0}^{t} \lambda X_{s} ds \label{DBOUWdefn}
\end{align}
for all $t \in \mathbb{R}$. Since $X$ is an OU process, we observe that the process $(W_{t})_{t\ge 0}$ is a standard Brownian motion\footnote{It will transpire that we are only interested in the increments of $W$; thus it does not matter that $W_{0}$ is known before time 0.}.  

\subsection{The dividend process}

We now define the dividend process by:
\begin{align*} 
\delta_{t} = a_{0} + a_{1} X_{t} + a_{2} X_{t}^{2}
\end{align*} for some constants $a_{0}, a_{1}, a_{2}$.  

The simplest non-trivial setup is that in which $a_{0}=a_{2}=0$, in which case the dividend process will simply be an OU process.  However, choosing such values of $a_{0}$ and $a_{2}$ means that there is a positive probability that the dividend process will become negative, which is unrealistic.  To overcome this problem, the constants can be chosen so that $a_{0} \ge a_{1}^{2}/4 a_{2}$, in which case the dividend process will always be non-negative.  Furthermore, it will transpire that considering the case in which the dividend process is a quadratic function of $X$ is no more difficult than the case in which $\delta$ is simply a scaling of $X$.\footnote{The case in which $\delta$ is a quadratic function of $X$ is slightly more complicated, since two different values of $X$ can give the same value of $\delta$. Hence, $\sigma(X_{s}:t_{0} \le s \le t) \ne \sigma(\delta_{s}:t_{0} \le s \le t)$.  Thus, we must assume that the agents observe the process $X$, rather than just observing the process $\delta$. }

\subsection{The Agents}

In our model there are $N$ agents at all times.  We assume that each person has a random lifetime.  When this person dies, their wealth is immediately passed onto their (ignorant) child.  Thus we are viewing each agent as a dynasty rather than a person\footnote{This idea of dynasties has been used by \citet{Nakata}, who considers an economy in which at any time point there are $H$ young and $H$ old agents. Each agent lives for 2 periods.   Young agent $h \in \{1, ... , H\}$ has the same preferences and beliefs as the old agent $h$. He then considers a Rational Beliefs Equilibrium as explained by Kurz.  However, all agents in his model live for exactly two units of time, in contrast to our assumptions.}.

Formally, there exist times $(T_{k}^{i})_{k \in \mathbb{Z}}$ which are the jump times of a stationary renewal process.  At each of these times $T_{k}^{i}$, agent $i$ will die and be replaced by his child.  Thus, the wealth of the agent will be maintained, but their beliefs will not;  the child will start with his own ignorant beliefs which will not depend on any historical data.   

Turning now to the beliefs of the agents, first recall that, under the reference measure,  $(X_{t})_{t\in\mathbb{R}}$ is an OU process with zero mean.  However, under the true measure, $X$ will revert to level $a$, which will not necessarily be zero.  The agents do not know this level.  They will use Bayesian updating to deduce it. 

We need to determine the measure that each agent works under.  First note that if we restrict to the time interval $[s,t]$, we may define a new measure by:
\begin{align}
\frac{d\mathbb{P}_{a}}{d\mathbb{P}_{0}} = \exp \left(\lambda a (W_{t}-W_{s}) - \frac{1}{2} (\lambda a)^{2} (t-s) \right)  \label{DBOUChangeOfMeasure1}
\end{align}
It follows from the Cameron-Martin-Girsanov theorem\footnote{See \citet{RogersWilliams}, IV.38 for an account} that a standard Brownian motion under $\mathbb{P}_{0}$ becomes a Brownian motion with drift $\lambda a$ under $\mathbb{P}_{a}$.  Formally,  $W_{r}=\bar{W}_{r}+\lambda a r$, for $s \le r \le t$ where $\bar{W}$ is a standard Brownian motion under $\mathbb{P}_{a}$.  Thus, 
\begin{align*}
dX_{t}= d\bar{W}_{t} + \lambda(a - X_{t}) dt
\end{align*}
so we see that, under $\mathbb{P}_{a}$, $X$ is an OU process which reverts to mean $a$.

Since agents do not know $a$, the beliefs of each agent simply consist of their distribution function for the parameter $a$.  When a member of the $i$th dynasty is born, he gives $ \lambda a$ a prior distribution\footnote{This is equivalent to having a prior distribution for $a$, since $\lambda$ is known.}.  We make the reasonable modelling assumption that this child's prior for $\lambda a$ is Normal  with mean $\alpha_{i}$ and precision\footnote{Equivalently, the prior has variance $\epsilon^{-1}$} \footnote{Hence, all members of dynasty $i$ begin life with the same prior} $\epsilon$.  The agent then updates his prior according to his observation of $(X_{s})_{ t_{k}^{i} \le s \le t}$, where $t_{k}^{i}$ denotes the time of birth of the current child and $t$ is the current time. 

If the agent knew the value of $a$, he would simply use a change of measure of the form (\ref{DBOUChangeOfMeasure1}).   However, $a$ is unknown, so the agent must weight each of the changes of measure according to his prior distribution for $a$.   Hence at time $t$, agent $i$'s law for the path has density with respect to the reference measure given by:
\begin{align}
\Lambda_{t}^{i}  &= \int_{-\infty}^{\infty} \sqrt{\frac{\epsilon}{2\pi}} \exp \Big(-\frac{\epsilon}{2}(\lambda a - \alpha^{i})^{2} + \lambda a (W_{t} - W_{t_{k}^{i}} ) - \frac{1}{2} (\lambda a)^{2} (t-t_{k}^{i}) \Big) d (\lambda a) \notag \\
 & = \sqrt{\frac{\epsilon}{\epsilon+\Delta t}} \exp \left(\frac{(\Delta W)^{2} + 2 \alpha^{i} \epsilon \Delta W - \epsilon (\alpha^{i})^{2} \Delta t}{2(\epsilon + \Delta t)} \right) \label{DBOUlambdaitb}
\end{align}
where $\Delta t$ and $\Delta W$ are given by:
\begin{align}
\Delta t & \equiv t - t_{k}^{i}  \qquad \Delta W \equiv W_{t} - W_{ t_{k}^{i}} \qquad \alpha \equiv \lambda a  \label{DBOUDeltas}
\end{align}

\subsection{Deriving the State Price Density}

Associated with agent (or dynasty) $i$ is a utility function, which we take to be CARA: $U_{i}(t,x) = -\frac{1}{\gamma_{i}} e^{-\gamma_{i} x} e^{-\rho t}$. Here, $\rho$ is the discount factor, assumed to be the same for all agents.  The agents seek to maximise the expected  discounted utility of their consumption.  Thus, agent $i$'s objective is:
\begin{align}
max \quad \mathbb{E}_{0} \Big[\int_{t_{0}}^{\infty}  U_{i}(t, c_{t}^{i}) \Lambda^{i}_{t}  \Big]  \label{DBOUObjective1}
\end{align} 
where $t_{0}$ is some start value, which we will later allow to go to $-\infty$.  $\Lambda^{i}_{t}$ is the density derived in (\ref{DBOUlambdaitb}), which jumps at each of the times $T^{i}_{k}$.  

The objectives of the agents have the same form as the previous \citet{BrownRogersDiverseBeliefs}, so its theory can be used to derive a state price density.  In particular, by looking at the price of an arbitrary contingent claim we can deduce that:   
\begin{align*}
\zeta_{s} \nu_{i} =  U_{i}'(s, c_{s}^{i}) \Lambda^{i}_{s}
\end{align*}
where $\nu_{i}$ is some $\mathcal{F}_{t_{0}}$ random variable\footnote{We will shortly let $t_{0}$ tend to negative infinity and when this occurs, the $\mathcal{F}_{t_{0}}$ will be trivial, thus $\nu_{i}$ will just be a constant} \footnote{ $U'$ denotes the derivative of $U$ with respect to its second argument}. Recalling our expression for $U_{i}$ and taking logs, we obtain:
\begin{align}
\frac{\log \zeta_{t}}{\gamma_{i}} + \frac{\log \nu_{i}}{\gamma_{i}} = -\frac{\rho t}{\gamma_{i}} - c_{t}^{i} + \frac{\log \Lambda^{i}_{t}}{\gamma_{i}}  \label{DBOUlogZetaEqn1}
\end{align}
Summing (\ref{DBOUlogZetaEqn1}) over $i$ and using market clearing gives:
\begin{align*}
\log \zeta_{t}\frac{1}{N}\sum \frac{1}{\gamma_{i}} + \frac{1}{N} \sum \frac{\log \nu_{i}}{\gamma_{i}} = -\frac{1}{N}\sum \frac{\rho t}{\gamma_{i}} - \frac{\delta_{t}}{N} + \frac{1}{N} \sum \frac{\log \Lambda^{i}_{t}}{\gamma_{i}}  
\end{align*}

\subsection{A continuum of agents}\label{DBOUContinuumAgents}

Recall that there are $N$ different agents in our model.  We will now let $N$ tend to infinity so that we can examine the case in which there is  a continuum of agents.  We assume that $\frac{1}{N}\sum \frac{1}{\gamma_{i}}$ has a finite limit and denote this limit by:
\begin{align*}
\Gamma^{-1} \equiv \lim_{N} \frac{1}{N}\sum \frac{1}{\gamma_{i}}
\end{align*}
Abusing notation slightly, we use $a_{i}$ to denote the $\lim_{N \rightarrow \infty} \frac{a_{i}}{N}$.  Hence:
\begin{align}
\log \zeta_{t} + G' =  -\rho t - \Gamma (a_{1} X_{t} + a_{2} X_{t}^{2}) + \Gamma \lim_{N \rightarrow \infty} \sum \frac{1}{N \gamma_{i}} \log \Lambda^{i}_{t} \label{DBOUlogZetatEq3}
\end{align}
where $G'$ is some $\mathcal{F}_{t_{0}}$-measurable function.  We now let $t_{0}$ tend to negative infinity;  $\mathcal{F}_{t_{0}}$ then becomes trivial, so $G'$ becomes a simple constant\footnote{We note that as $t_{0} \rightarrow \infty$, the expression on the right of (\ref{DBOUlogZetatEq3}) is almost surely finite, so the left hand side must be as well.  Since our $\zeta$ and $(\nu_{i})_{1 \le i \le N}$ were only chosen up to a multiplicative constant, we may choose them to depend on $t_{0}$ in such a way that as $t_{0} \rightarrow \infty$ both $\zeta$ and $G'$ are a.s. finite. }.  

Only the last term in (\ref{DBOUlogZetatEq3}) requires further development.  Writing $u^{i}$ for the time since the the last person died in the $i$th dynasty, we obtain:
\begin{multline}
\Gamma \lim_{N \rightarrow \infty} \sum \frac{1}{N \gamma_{i}} \log \Lambda^{i}_{t} = \Gamma \lim_{N \rightarrow \infty} \sum \frac{1}{N \gamma_{i}} \Big[ \frac{1}{2}\log \Big(\frac{\epsilon}{\epsilon+u^{i}} \Big)  \\
+ \Big(\frac{(W_{t}-W_{t-u^{i}})^{2} + 2 \alpha^{i} \epsilon (W_{t}-W_{t-u^{i}}) - \epsilon (\alpha^{i})^{2} u^{i}}{2(\epsilon + u^{i})} \Big) \Big]  \label{DBOUCaseLogZeta11}
\end{multline}
We assume that the mean of $\alpha^{i}$ is given by $\langle \alpha \rangle$ and further that the distribution of $u^{i}, \alpha^{i}$ and $\gamma_{i}$ are all independent.  We further make the assumption that $u$ has a density $\varphi(\cdot)$, given by:
\begin{align*}
\varphi(u) = A (\epsilon + u) \lambda e^{-\lambda u}
\end{align*}
where $A= \frac{\lambda}{1+ \epsilon \lambda}$ is chosen so that $\int_{0}^{\infty} \varphi(u) du =1$. Since $\varphi(u)$ represents the probability of someone who is currently alive having age $u$, it follows that $\varphi(\cdot)$ must be decreasing.   This gives the inequality $\lambda \epsilon \ge 1$.  

Using our expression for $\varphi$, equation (\ref{DBOUCaseLogZeta11}) becomes:
\begin{multline*}
\log \zeta_{t} = -G - \Gamma (a_{1} X_{t} + a_{2} X_{t}^{2}) - \rho  t+ \frac{1}{2} \int \frac{ \big(W_{t} - W_{t - u}\big)^{2}}{\epsilon + u} \varphi(u)du \\
+  \langle \alpha \rangle \epsilon \int \frac{ \big(W_{t} - W_{t - u}\big)}{\epsilon + u} \varphi(u) du   
\end{multline*}
where $G$ is some new constant.  This then gives us:
\begin{align*}
 \log \zeta_{t} = -G - \Gamma (a_{1} X_{t} + a_{2} X_{t}^{2}) - \rho  t+ \frac{A}{2} \eta_{t} +  \langle \alpha \rangle \epsilon A \xi_{t}
\end{align*}
where
\begin{align*}
\xi_{t} &=\int_{0}^{\infty} (W_{t} - W_{t-u}) \lambda e^{-\lambda u} du \\
\eta_{t} &=\int_{0}^{\infty} (W_{t}- W_{t-u})^{2} \lambda e^{-\lambda u} du
\end{align*}
By rearrangement and use of Fubini (see appendix), we are able to show that:
\begin{align*}
\xi_{t} & = X_{t} \\
\eta_{t} & = X_{t}^{2} +  e^{-\lambda t} \int_{-\infty}^{t} \lambda e^{\lambda s} X_{s}^{2} ds
\end{align*}
Our final expression for the state price density is then given by:
\begin{align}
\log \zeta_{t} &= - G - \Gamma ( a_{1} X_{t} + a_{2} X_{t}^{2}) - \rho t + \frac{A}{2} [(X_{t})^{2} + e^{-\lambda t} \int_{-\infty}^{t} \lambda e^{\lambda s} X_{s}^{2} ds  ] + \langle \alpha \rangle \epsilon A X_{t} \\
&= - G + B X_{t} + C X_{t}^{2} + U_{t} - \rho t \label{DBOUlogzetatfinal}
\end{align}
where:
\begin{align*}
B = \langle \alpha \rangle \epsilon A - \Gamma a_{1} \qquad C= \frac{A}{2} - \Gamma a_{2}
\end{align*}
and
\begin{align*}
U_{t} = \frac{1}{2} A e^{-\lambda t} \int_{-\infty}^{t} \lambda e^{\lambda s} X_{s}^{2} ds
\end{align*}

\section{Asset Prices}\label{DBOUCaseAssetPrices}

\subsection{The Interest Rate Process}

We will use our state price density to derive the interest rate process.  From It\^{o}'s formula, we have:
\begin{multline*}
\frac{d \zeta_{t}}{\zeta_{t}}  = (B + 2 C X_{t}) dW_{t} \\
+  \big(C + \frac{\lambda A}{2} X_{t}^{2} - \lambda U_{t} - \rho - B \lambda X_{t} - 2 \lambda C X_{t}^{2} + \frac{1}{2} ( B+2C X_{t})^{2} \big) dt 
\end{multline*}
\begin{align*}
 \dot{=} \big[(-\rho + C+ \frac{1}{2} B^{2}) + (-\lambda B + 2 C B) X_{t} + (- 2 \lambda C + \frac{\lambda A}{2} + 2 C^{2} ) X_{t}^{2} - \lambda U_{t}  \big] dt
\end{align*}
where the symbol $\dot{=}$ signifies that the two sides differ by a local martingale.  The interest rate is equal to minus the coefficient of $dt$ in the above expansion, hence:
\begin{align}
r_{t} =  r(X_{t}, U_{t}) \equiv (\rho - C - \frac{1}{2} B^{2}) + B (\lambda - 2 C) X_{t} + (2 \lambda C - \frac{\lambda A}{2} - 2 C^{2} ) X_{t}^{2} + \lambda U_{t} \label{DBOUInterestRate}
\end{align}
Thus, our model gives us an interest rate process of the form:
\begin{align*}
r_{t} = \alpha_{0} + \alpha_{1} X_{t} + \alpha_{2} X_{t}^{2} + \lambda U_{t}
\end{align*}
for some constants $\alpha_{i}, i=0,1,2$.  Note that the interest rate process will depend on the behaviour of the dividend process in the past (via $U_{t}$) as well as on the current value of the dividend process.  We therefore see that in some sense, high historical volatility generates high values of the riskless rate.

\subsection{The Stock Price}

We will now calculate the stock price.  We have:   
\begin{align}
S_{t} & = \mathbb{E}_{t}^{0} \Big[\int_{t}^{\infty} \frac{\zeta_{u} \delta_{u}}{\zeta_{t}} du  \Big] \notag \\
& = \frac{1}{\zeta_{t}} \int_{t}^{\infty} \mathbb{E}_{t}^{0} \Big[\zeta_{u} \delta_{u}  \Big] du  \label{DBOUStockPriceIntegral}
\end{align}

\subsubsection{A PDE for the stock price}

From the form of $\zeta_{t}$ and the Markovian structure, we will have that:
\begin{align}
\zeta_{t} S_{t} = \zeta_{t} h(X_{t}, U_{t}) \label{DBOUdefh}
\end{align}
for some function $h$.  This function will satisfy a PDE which we may determine by by observing that $\zeta_{t} S_{t} + \int_{0}^{t} \zeta_{s} \delta_{s} ds$ is a martingale and applying It\^{o}'s formula.   After a few calculations, we obtain the PDE:
\begin{align}
0 = \frac{1}{2} h_{xx} + (B + (2 C - \lambda) x) h_{x} + \lambda (\frac{A}{2} x^{2} - u) h_{u} - r(x,u) h + (a_{0} + a_{1} x + a_{2} x^{2}) \label{DBOUhPDE}
\end{align}
Unfortunately, it does not appear to be possible to solve this equation in closed form, so we will resort to another approach.  However, before we do this, let us look at some of the consequences of (\ref{DBOUdefh}) and (\ref{DBOUhPDE}).  Suppose that under the real-world probability, $\mathbb{P}_{*}$, the OU process reverts to level $a^{*}$, then we have that:
\begin{align*}
dS_{t} = h_{x} dW_{t}^{*} + h_{x} \lambda (a^{*} - X_{t}) dt + h_{u} (\frac{\lambda A}{2} X_{t}^{2} - \lambda U_{t}) dt + \frac{1}{2} h_{xx} dt 
\end{align*}
where $W^{*}$ denotes a Brownian motion under measure $\mathbb{P}^{*}$.  After using (\ref{DBOUhPDE}) we get that:
\begin{align*}
dS_{t} = h_{x} dW_{t}^{*} + h_{x} \big( (\lambda a^{*} - B) - 2 C X_{t}  \big) dt + r(X_{t}, U_{t}) h dt - (a_{0} + a_{1} X_{t} + a_{2} X_{t}^{2}) dt
\end{align*}
Hence, we see that the volatility and drift of the stock price are given by:
\begin{align}
\Sigma_{t} & = \frac{h_{x}(X_{t}, U_{t})}{h(X_{t}, U_{t})} \label{DBOUVolatilityFromPDE}\\
\mu_{t}^{*} & = \frac{r(X_{t}, U_{t})h(X_{t}, U_{t}) - (a_{0} + a_{1} X_{t} + a_{2} X_{t}^{2}) + \big(\lambda a^{*} - 2 C X_{t} - B \big) h_{x}(X_{t},U_{t})}{h(X_{t}, U_{t})} \label{DBOUDriftFromPDE}
\end{align}
We shall use these expressions later.  

\subsubsection{Calculation of stock price via computation of conditional expectation}

We will now proceed to determine the stock price via another method.  Substituting the state price density from (\ref{DBOUlogzetatfinal}) into (\ref{DBOUStockPriceIntegral}), we obtain:
\begin{multline*}
S_{t} = \exp\{-B X_{t} - C X_{t}^{2} - U_{t} + \rho t  \}  \\
\int_{t}^{\infty} \mathbb{E}_{t}^{0} \big[(a_{0} + a_{1} X_{T} + a_{2} X_{T}^{2}) \exp \{B X_{T} + C X_{T}^{2} + U_{T} - \rho T  \}  \big] dT
\end{multline*}
On first sight it may appear that it is very difficult to get any further with this expression.  However, if we can calculate:
\begin{align*}
V^{T}(t, X_{t}; \theta) := \mathbb{E}_{t}^{0} \big[ \exp \{\theta (a_{0} + a_{1} X_{T} + a_{2} X_{T}^{2}) + B X_{T} + C X_{T}^{2} + \int_{t}^{T} \frac{A}{2} \lambda e^{\lambda (s-T)} X_{s}^{2} ds  \} \big]
\end{align*}
then we may differentiate with respect to $\theta$ to and set $\theta=0$ to give:
\begin{align*}
S_{t} = \exp \{ -B X_{t} - C X_{t}^{2} \}  \int_{t}^{\infty} \exp \{(e^{-\lambda (T-t)} - 1) U_{t} - \rho (T-t)\} \frac{\partial}{\partial \theta}|_{\theta=0} V^{T}(t, X_{t}; \theta) dT
\end{align*}
We also define $\tau \equiv T-t$.  We will show that:
\begin{align*}
V^{T}(t, X_{t}; \theta) = \exp \{ \frac{1}{2} a(\tau) X_{t}^{2} + b(\tau) X_{t} + c(\tau) \}
\end{align*}
where $a,b$ and $c$ are functions which we will shortly deduce.  To deduce these functions, we will use a martingale argument.  For $t \le T$ we  define:
\begin{align*}
M_{t}^{T} & \equiv \mathbb{E}_{t}^{0} \big[\exp \{\theta (a_{0} + a_{1} X_{T} + a_{2} X_{T}^{2}) + B X_{T} + C X_{T}^{2} + \int_{-\infty}^{T} \frac{A}{2} \lambda e^{\lambda (s-T)} X_{s}^{2} ds  \}  \big] \\ 
& = V^{T}(t, X_{t}; \theta) \exp \{ \int_{-\infty}^{t} \frac{A}{2} \lambda e^{\lambda (s-T)} X_{s}^{2} ds \}
\end{align*}
Now apply It\^{o}'s formula:
\begin{align*}
dM_{t}^{T}  = \exp \{ \int_{-\infty}^{t} \frac{A}{2} \lambda e^{\lambda (s-T)} X_{s}^{2} ds \} \big[ V_{t} dt + V_{x} dX_{t} + \frac{1}{2} V_{xx} dX_{t} dX_{t} + \frac{\lambda A}{2} e^{\lambda(t-T)} X_{t}^{2} V dt  \big] 
\end{align*}
\begin{multline*}
 = M_{t}^{T} \big[\frac{\lambda A}{2} e^{\lambda(t-T)} X_{t}^{2} dt  - (\frac{1}{2} \dot{a}(\tau) X_{t}^{2} + \dot{b}(\tau) X_{t} + \dot{c}(\tau) ) dt  \\
+ (a(\tau) X_{t} + b(\tau))( dW_{t} - \lambda X_{t} dt) + \frac{1}{2} (a(\tau) + (a(\tau) X_{t} + b(\tau))^{2} )dt \big]
\end{multline*}
But $(M_{t}^{T})_{t \le T}$ is a martingale under $\mathbb{P}_{0}$, so the coefficient of $dt$ in the above expression must be zero.  Thus we obtain:
\begin{align*}
\frac{1}{2} \dot{a} &= \frac{\lambda A}{2} e^{-\lambda \tau} - \lambda a + \frac{1}{2} a^{2} \\
\dot{b} &= ab - \lambda b \\
\dot{c} &= \frac{1}{2} (a + b^{2})
\end{align*}
The boundary conditions are given by:
\begin{align*}
a(0) = 2 (C + \theta a_{2}) \qquad b(0) = B + \theta a_{1} \qquad c(0)= \theta a_{0}
\end{align*}

\subsubsection{Solving the ODEs}

We now solve the ODEs.  The first equation is a Riccati equation, so in order to solve we make the usual substitution:
\begin{align*}
a(\tau) = - \frac{ \dot{g} (\tau)}{g(\tau)}
\end{align*}
Substituting this into the ODE for $a$ gives:
\begin{align*}
\frac{1}{2} \ddot{g} + \lambda \dot{g} + \frac{\lambda A}{2} e^{-\lambda \tau} g = 0
\end{align*}
and the boundary condition becomes:
\begin{align*}
- \dot{g}(0)= 2(C+ \theta a_{2}) g(0)
\end{align*}
We can solve this equation using Maple to obtain:
\begin{multline*}
g(u)= e^{-\lambda u} \Big[\big(\sqrt{\lambda A} Y_{1} (2 \sqrt{A/\lambda}) - 2(C+ \theta a_{2}) Y_{2}(2 \sqrt{A/\lambda})  \big) J_{2}(2 e^{-\lambda u/2} \sqrt{A/\lambda})  \\  -\big(\sqrt{\lambda A} J_{1} (2 \sqrt{A/\lambda}) - 2(C+ \theta a_{2}) J_{2}(2 \sqrt{A/\lambda})  \big) Y_{2}(2 e^{-\lambda u/2} \sqrt{A/\lambda}) \Big]
\end{multline*}
where $J_{i}$ and $Y_{i}$ are Bessel functions of order $i$ of the first and second kind respectively.  Turning now to the ODE for $b$, we may use our solution for $a$ to deduce:
\begin{align*}
\dot{b} + \frac{\dot{g}}{g} b + \lambda b = 0
\end{align*}
Rearranging gives:
\begin{align*}
\frac{d}{d\tau} ( b g e^{\lambda \tau}) = 0 
\end{align*}
which we can solve subject to $b(0)= B+ \theta a_{1}$ to give:
\begin{align*}
b(\tau) = \frac{(B+\theta a_{1}) g(0)}{e^{\lambda \tau} g(\tau)}
\end{align*}
Finally, we obtain:
\begin{align*}
c(\tau) = \theta a_{0} + \int_{0}^{\tau} \frac{1}{2} (a(\tau') + b(\tau')^{2} ) d\tau'
\end{align*}
Thus we have completely solved the ODEs.  In order to calculate the stock price, we need to find $\frac{\partial V}{\partial \theta}$.  We therefore need:
\begin{align*}
\frac{\partial g }{\partial \theta} = e^{-\lambda u} \big[ -2 a_{2} Y_{2} (2 \sqrt{A/\lambda}) J_{2}(2 e^{-\lambda u/2} \sqrt{A/\lambda}) + 2 a_{2} J_{2} (2 \sqrt{A/\lambda}) Y_{2}(2 e^{-\lambda u/2} \sqrt{A/\lambda})  \big]
\end{align*}
and also:
\begin{multline*}
\frac{\partial \dot{g}}{\partial \theta} = - \lambda \frac{\partial g }{\partial \theta} + e^{-\lambda u} \big[-2a_{2} Y_{2} (2 \sqrt{A/\lambda}) \big( \lambda J_{2} (2 \sqrt{A/\lambda} e^{-\lambda u/2}) - \sqrt{A \lambda} e^{-\lambda u /2} J_{1} (2 \sqrt{A/\lambda} e^{-\lambda u/2})\big)  \\
+ 2a_{2} J_{2} (2 \sqrt{A/\lambda}) \big( \lambda Y_{2} (2 \sqrt{A/\lambda} e^{-\lambda u/2}) - \sqrt{A \lambda} e^{-\lambda u /2} Y_{1} (2 \sqrt{A/\lambda} e^{-\lambda u/2})\big) \big]
\end{multline*}
We may then calculate expressions for $\frac{\partial V}{\partial \theta}$.  First note that:
\begin{align*}
\frac{\partial V}{\partial \theta} = \big(\frac{1}{2} \frac{\partial a}{\partial \theta} X_{t}^{2} + \frac{\partial b}{\partial \theta} X_{t} + \frac{\partial c}{\partial \theta}\big)  \exp \{ \frac{1}{2} a(\tau) X_{t}^{2} + b(\tau) X_{t} + c(\tau) \}
\end{align*}
But:
\begin{align*}
\frac{\partial c}{\partial \theta}(\tau) &= a_{0} + \int_{0}^{\tau} \frac{1}{2} \big(\frac{\partial a}{\partial \theta}(\tau') + 2 b(\tau') \frac{\partial b}{\partial \theta}(\tau')  \big) d \tau' \\
\frac{\partial b}{\partial \theta}(\tau)  &= a_{1} \frac{g(0)}{e^{\lambda \tau} g(\tau)} + \frac{(B+\theta a_{1})}{e^{\lambda \tau}} \frac{ \frac{\partial g}{\partial \theta}(0)}{g(\tau)} - \frac{(B+\theta a_{1})}{e^{\lambda \tau}} \frac{g(0)}{g(\tau)^{2}} \frac{\partial g}{\partial \theta}(\tau) \\
\frac{\partial a}{\partial \theta}(\tau) &= - \frac{\frac{\partial \dot{g}}{\partial \theta}(\tau)}{g(\tau)} + \frac{\dot{g(\tau)}}{g(\tau)^{2}} \frac{\partial g}{\partial \theta}(\tau)
\end{align*}
So finally we have:
\begin{multline}
S_{t} = \exp \{- B X_{t} - C X_{t}^{2} \} \int_{0}^{\infty} \exp \{ - \rho \tau - (1-e^{-\lambda \tau}) U_{t} \} \\
\big(\frac{1}{2} \frac{\partial a}{\partial \theta} X_{t}^{2} + \frac{\partial b}{\partial \theta} X_{t} + \frac{\partial c}{\partial \theta}\big)  \exp \{ \frac{1}{2} a(\tau) X_{t}^{2} + b(\tau) X_{t} + c(\tau) \}  d \tau \label{DBOUStockPriceFormula}
\end{multline}
This is as far as we can get with the expression for the stock price.  We see that the stock price depends not only on the dividend at time $t$, but also on $U_{t}$, a term reflecting the behaviour of $(X_{s})_{-\infty \le s \le t}$.  This is as we would expect, since agents need to use information from the whole of their lifetimes to make better estimates of the mean to which $X$ is reverting. From properties of the OU process, we see that if $X_{t}$ reverts to mean $a$ then, since $X$ is stationary, we have $X_{t} \sim N(a, \frac{1}{2\lambda})$.  Hence, 
\begin{align*}
\mathbb{E} U_{t} = \int_{-\infty}^{t} \frac{\lambda A}{2} e^{\lambda (r-t)} \big(\frac{1}{2\lambda} + a^{2} \big) dr = \frac{A}{2} \big(\frac{1}{2\lambda} + a^{2}  \big)
\end{align*}  
This indicates a sensible value for $U_{t}$, which will be helpful for when we begin to look at numerical examples later on.

\subsection{The Bond Price}

The time-$t$ price of a zero-coupon bond which has unit payoff at time $T$ is given by:
\begin{align*}
\mathbb{E}^{0} \Big[\frac{\zeta_{T}}{\zeta_{t}} | \mathcal{F}_{t} \Big]= \exp \Big[ - B X_{t} - C X_{t}^{2} - U_{t} (1- e^{-\lambda \tau}) - \rho \tau  \Big]V^{T}(t,X_{t};\theta=0)
\end{align*}
Using our expression for $V^{T}(t,X_{t}; \theta=0)$, we obtain:
\begin{multline}
\exp \Big[ (\frac{1}{2} a(\tau)-C) X_{t}^{2} + (b(\tau)-B) X_{t} + c(\tau) - \rho \tau - (1-e^{-\lambda \tau}) U_{t} \Big] \label{DBOUBondPrice}
\end{multline}
where the functions $a,b$ and $c$ are all evaluated using $\theta=0$. 

\subsection{Remarks on the case in which $a$ is known}

Note that if we let $\epsilon \rightarrow \infty$, then this corresponds to the case in which all the agents are certain that they know the value of $a$.  By taking the limit in our expressions for the stock price, bond price and riskless rate, we can deduce expressions for these quantities in this limit.  We note further that if the agents are sure about the value of $a$ and this value corresponds to the true value, $a^{*}$, then the expressions we obtain will be the same as those for the model in which the true value of $a$ was known to all the agents.

\section{Numerical Results}\label{DBOUCaseNumRes}

\subsection{Calibrating the model}

We now investigate the qualitative behaviour of the stock price as we vary the parameters in our model. To do this it is important that we choose a suitable region over which the parameters can vary.  We will restrict to the case in which $a_{0}=a_{1}=0$, so that we have simply $\delta_{t} = a_{2} X_{t}^{2}$.  This ensures that the dividend process remains positive.  Note further that the state price density (\ref{DBOUlogzetatfinal}) only depends on the \emph{product} $\Gamma a_{2}$ rather than the individual $\Gamma$ and $a_{2}$.  Although the dividend process does depend on $a_{2}$, changing $a_{2}$ simply corresponds to the changing the units in which we measure the dividend process.  Hence, we may choose $a_{2}=1$. 

Some of the parameters are relatively easy to choose, such as $\lambda$ and $\rho$, for which we choose $\lambda=2$ and $\rho=0.04$. However, other parameters, such as $\Gamma$ are much harder to determine.  We are only interested in ensuring that the parameters are of the correct order.  In order to do this, we will begin by assuming that $\langle \alpha \rangle = a$.  Furthermore, for the purposes of calibration, we will consider the limit as $\epsilon \rightarrow \infty$, which corresponds to the case in which agents are sure that they know the true value of $a$.  This leaves the parameters $a$ and $\Gamma$ which we still need to determine.  

One way to determine these parameters would be to choose them in order to match various moments from empirical data, such as the mean price-dividend ratio; this was the strategy employed in \citet{BrownRogersDiverseBeliefs} when we considered the equity premium puzzle.  Ideally, we would use the same method here, but unfortunately our stock price is much more complicated.  Thus, computing a given stock price requires the numerical computation of an integral.  To work out the mean price dividend ratio, we would then need to compute a further integral as we averaged over the values of the driving Brownian motion.  We would then vary the parameters and calculate the expected price dividend ratio each time in an attempt to find a realistic set of parameters.  Given the additional complexity of this problem and the fact that we are only interested in determining parameters that are of the correct order, we will proceed in a different manner.

We first note that the interest rate process has a particularly simple form, which we can use to get a simple expression for the expected riskless rate. We can match this with the mean riskless rate from the Shiller data set.  

Note that we are considering the case in which $a_{0}=a_{1}=0, a_{2}=1$ and the limit as $\epsilon \rightarrow \infty$ and hence $A \epsilon \rightarrow 1, B \rightarrow \langle \alpha \rangle = a, C \rightarrow  - \Gamma$.  Substituting into expression (\ref{DBOUInterestRate}) gives:
\begin{align*}
r = (\rho + \Gamma - \tfrac{1}{2} a^{2}) + a(\lambda + 2 \Gamma ) X_{t} - 2 \Gamma (\lambda+ \Gamma ) X_{t}^{2} 
\end{align*}
Thus, the expected riskless rate is given by:
\begin{align*}
\mathbb{E} r = (\rho + \Gamma - \tfrac{1}{2} a^{2}) + a^{2} (\lambda + 2 \Gamma ) - 2 \Gamma (\lambda+ \Gamma )(a^{2} + \tfrac{1}{2\lambda})
\end{align*}

To determine $\Gamma$, we compare a CRRA agent (where we know a reasonable value for the constant of relative risk aversion) with a CARA agent.  If we consider a single agent model in which the value of $a$ is known, the stock price will be given by:
\begin{align*}
S_{0} = \mathbb{E} \int_{0}^{\infty} \frac{U'(\delta_{t})}{U'(\delta_{0})} \delta_{t} dt
\end{align*}
Since we just want our parameters to be of the correct order, it is sufficient to check that the behaviour of 
\begin{align}
\frac{U'(\delta_{t})}{U'(\delta_{0})} \delta_{t}  \label{DBOUUtilityRatio}
\end{align}
in the neighbourhood of $X=a$ is the same for both the CRRA and CARA case.  If we set $X_{0}=X_{t}=a$ then clearly (\ref{DBOUUtilityRatio}) will be the same in both the CRRA and CARA case.  We therefore impose the requirement that a small change in $X_{t}$ from $X_{t}=a$ has the same effect in both cases, leading to the condition:
\begin{align*}
\frac{U''_{CRRA}( a^{2})}{U'_{CRRA}( a^{2})}=\frac{U''_{CARA}( a^{2})}{U'_{CARA}( a^{2})}
\end{align*} 
Which leads us to the condition:
\begin{align}
\Gamma = \frac{R}{a^{2}}
\end{align}
Since we know a sensible value for $R$ is $R=2$, this equation gives us an equation from which we can determine $\Gamma$ and $a$.  Substituting in our expression for the riskless rate yields the cubic equation:
\begin{align*}
l(\Gamma) \equiv \frac{\Gamma^{3}}{\lambda} + 2 R \Gamma^{2} + (\mathbb{E} r - \rho + 2 R (\lambda - 1) ) \Gamma + \tfrac{1}{2} R -  R \lambda  =0
\end{align*}
We will choose $R=2$.  We also choose $\mathbb{E}r=0.01$, as given by the Shiller data set. We may then note that $l(0)<0$ and $\tfrac{dl}{d\Gamma}>0$ for $\Gamma>0$, hence there is a unique positive solution to the above equation, which we can easily compute.  Computation shows that the correct $\Gamma$ to choose is $\Gamma=0.49$ which we take as our default value.  This gives $a$=2.01.

To summarise, the default parameters we choose are: $a_{0}=a_{1}=0; a_{2}=1; \lambda=2; \rho=0.04; \epsilon=1.0; \Gamma=0.49; \langle \alpha \rangle = a=2.01$.  We also choose $X_{t}=a, U_{t}=\dfrac{A}{2}(a^{2} + \tfrac{1}{2\lambda})$.  We then vary the parameters and examine the behaviour.

\subsection{Comments on Results}

\begin{figure}[p]
  \caption{Graph of $S_{t}$ against $\lambda$. }
  \begin{center}
    \includegraphics[width=1.0\textwidth]{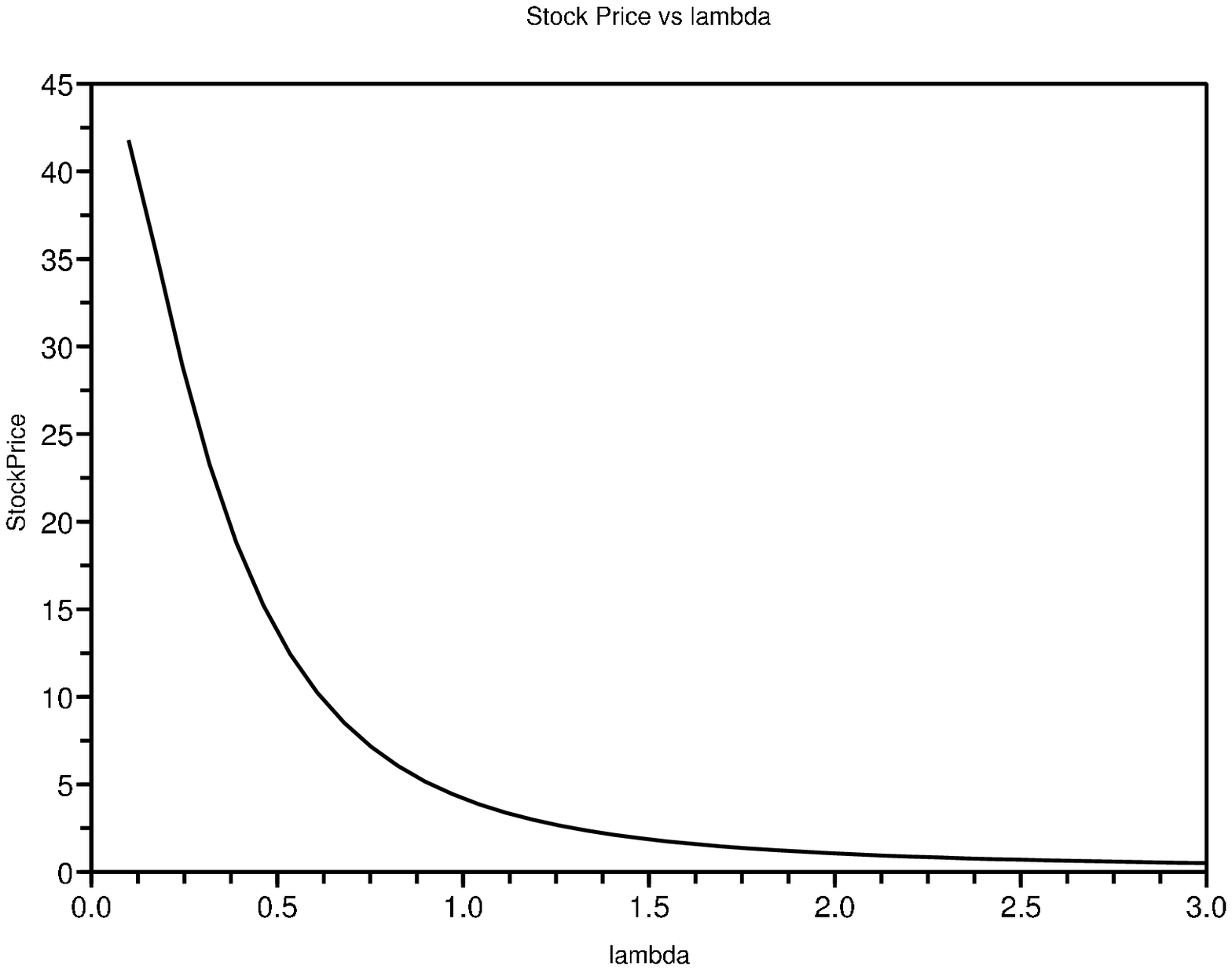}
  \end{center}
  \label{DBOUGraphStvslambda}
\end{figure}

Figure \ref{DBOUGraphStvslambda} shows that the stock price is decreasing in $\lambda$.   Recall that $\lambda$ is the parameter which tells us how quickly the dividend process returns to its mean.  Hence, a lower value of $\lambda$ means that the dividend process is more likely to reach high values, so is worth more to the agents.  However, $\lambda$ is also a parameter used in specifying the distribution of the lifetime of the agents. Increasing $\lambda$ therefore decreases the expected lifetime of the agents.  Each child in the dynasty therefore has less time to learn about the unknown parameter $a$ and this increased uncertainty amongst the agents also means that the stock price decreases as $\lambda$ increases.

\begin{figure}[p]
  \caption{Graph of $S_{t}$ against $\epsilon$. }
  \begin{center}
    \includegraphics[width=1.0\textwidth]{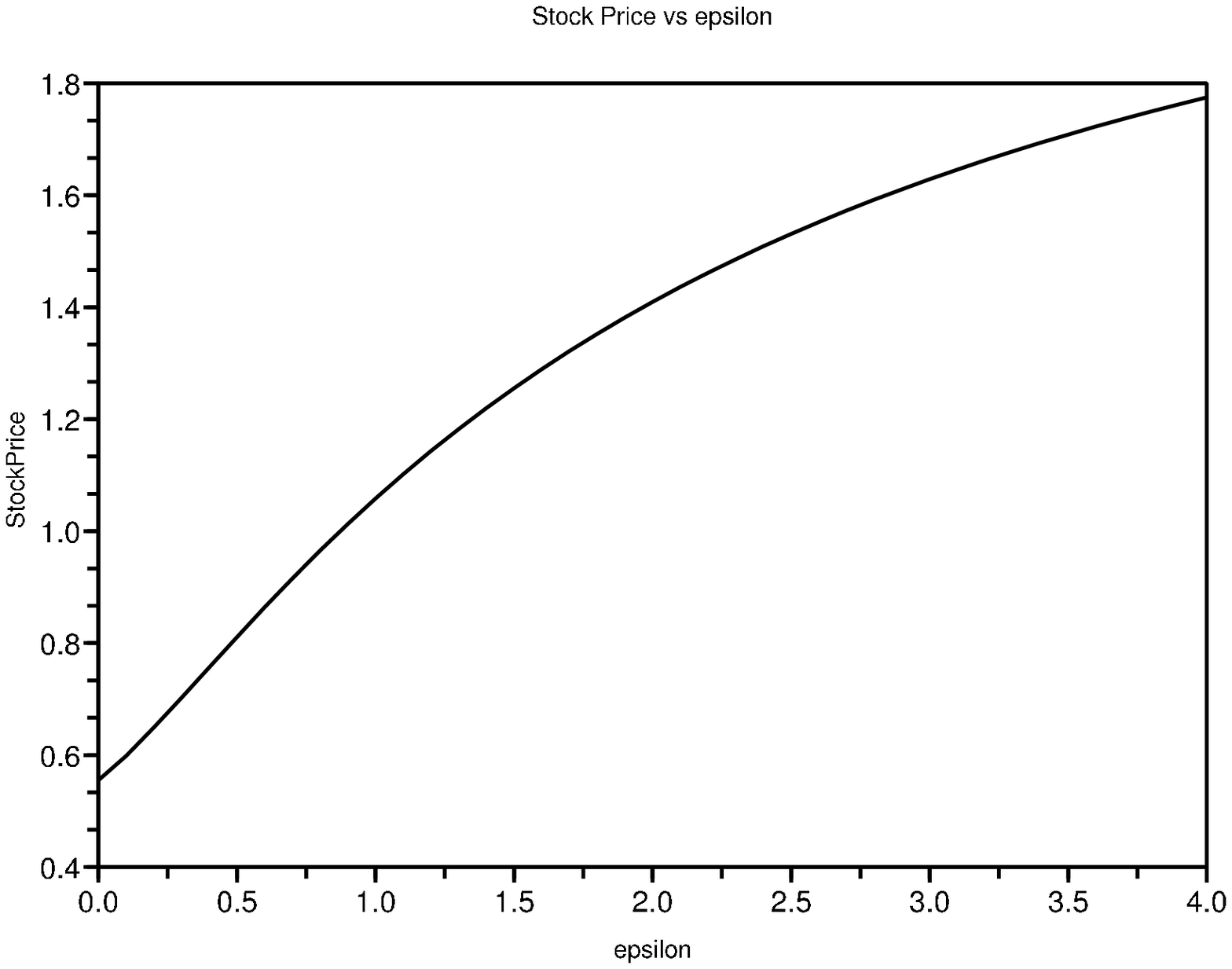}
  \end{center}
  \label{DBOUGraphStvseps}
\end{figure}

\begin{figure}[p]
  \caption{Graph of $S_{t}$ against $ \rho $. }
  \begin{center}
    \includegraphics[width=1.0\textwidth]{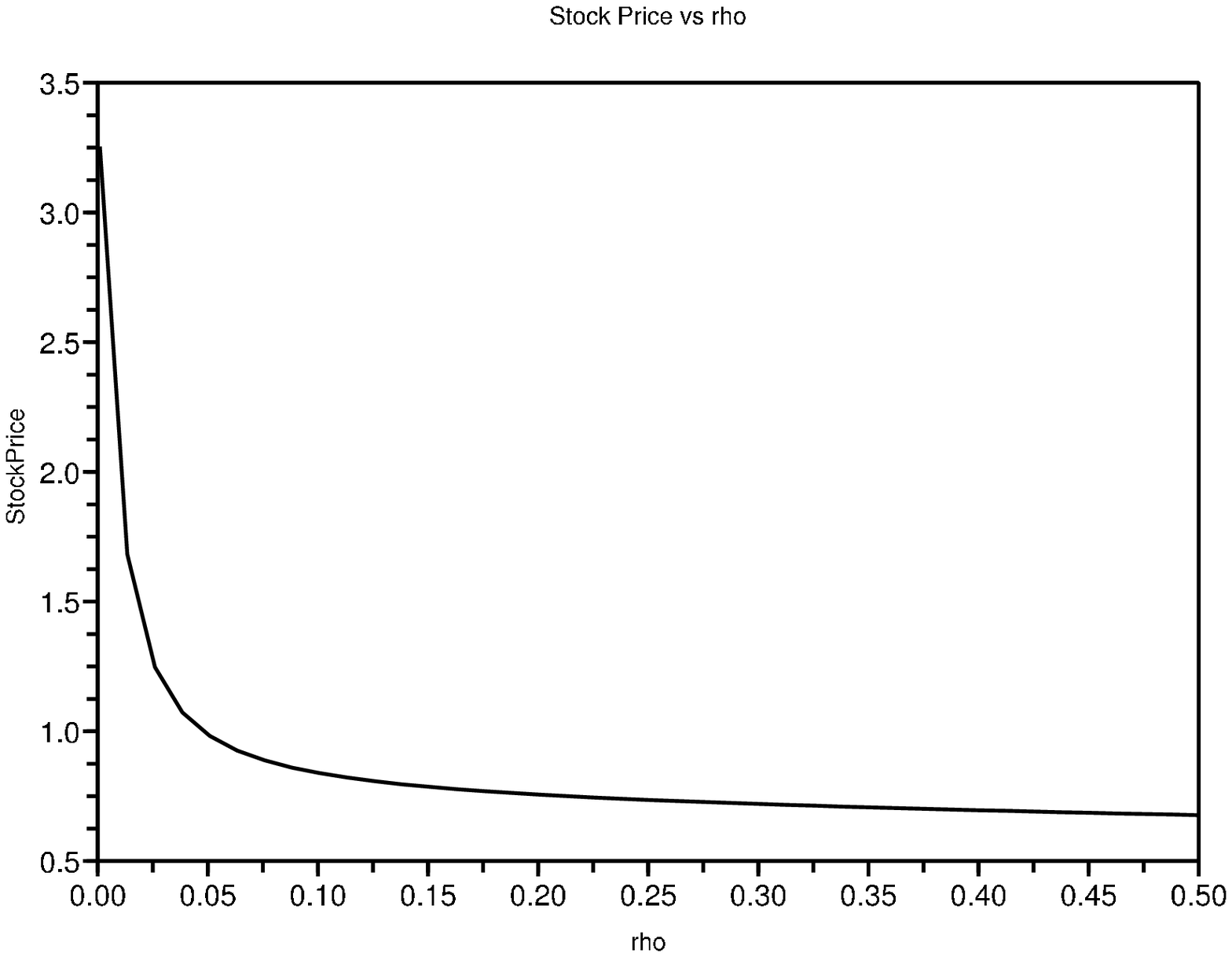}
  \end{center}
  \label{DBOUGraphStvsrho}
\end{figure}

Figure \ref{DBOUGraphStvseps} shows that as $\epsilon$ increases, so does the stock price, which is to be expected since if the agents know more about the dividend process (i.e. their beliefs have a higher precision), the stock should be worth more to them.  Similarly, Figure \ref{DBOUGraphStvsrho} shows that the larger the value of $\rho$, the less the stock is worth.  A large $\rho$ indicates that the agents are impatient and want to consume their wealth in the near future, making the stock less attractive.  

\begin{figure}[p]
  \caption{Graph of $S_{t}$ against $\langle \alpha \rangle$. }
  \begin{center}
    \includegraphics[width=1.0\textwidth]{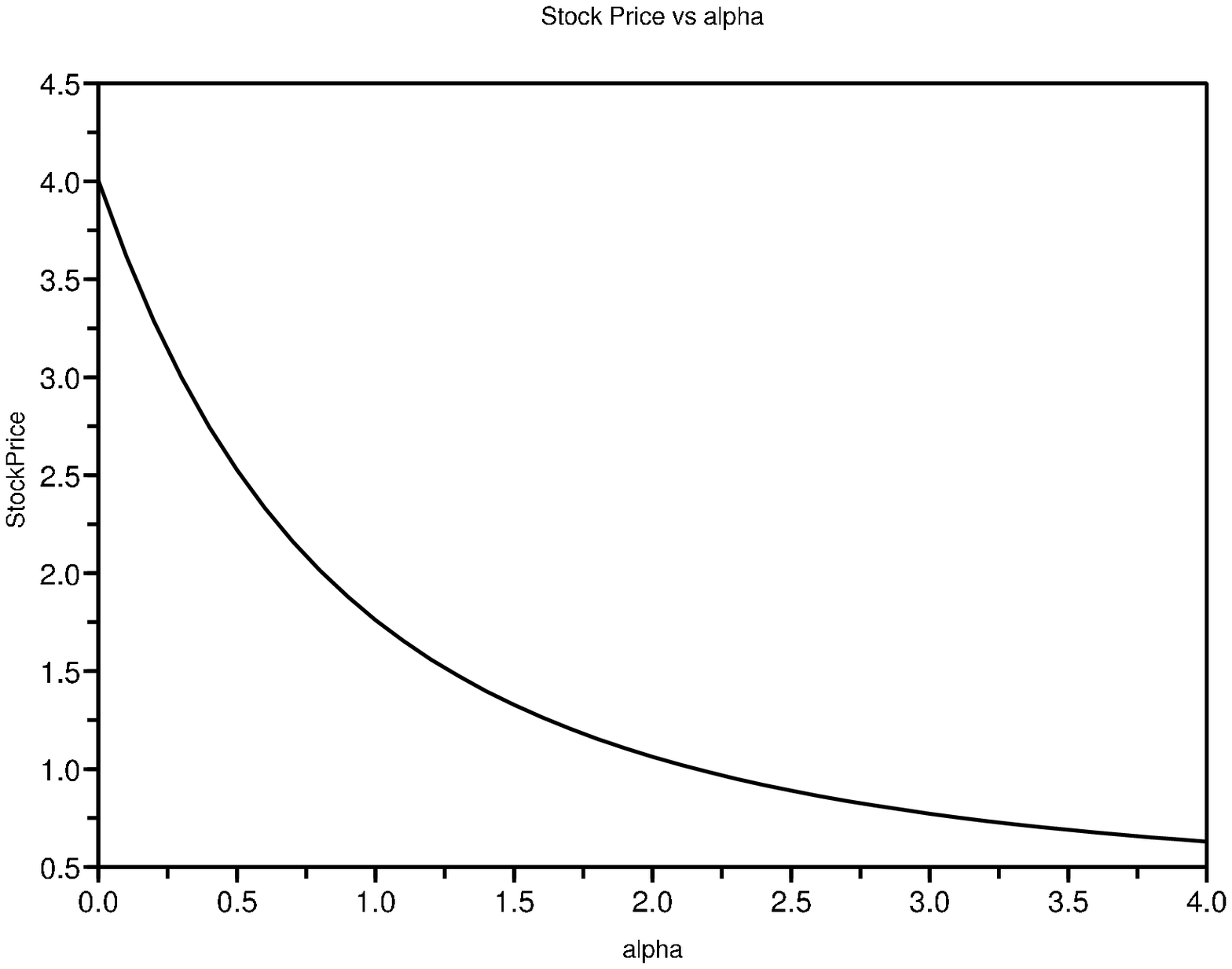}
  \end{center}
  \label{DBOUGraphStvsalpha}
\end{figure}

 Figure \ref{DBOUGraphStvsalpha} exhibits the dependence of the stock price on $\langle \alpha \rangle$.    Recall that $X_{t}$ and $U_{t}$ are kept fixed as we vary $\langle \alpha \rangle$.  A small $\langle \alpha \rangle$ indicates that the agents think the level to which $X$ reverts is low.  Thus, since we do not change $X_{t}$, a low value of $\langle \alpha \rangle$ relative to $X$ indicates that $X$ is currently abnormally high and so the dividends are abnormally high.  Thus, the agents are keen to hold this stock.  Furthermore, the relatively high level of $X$ means that the agents have a large amount of dividend with which to buy the stock.

\begin{figure}[p]
  \caption{Graph of $S_{t}$ against $\Gamma$. }
  \begin{center}
    \includegraphics[width=1.0\textwidth]{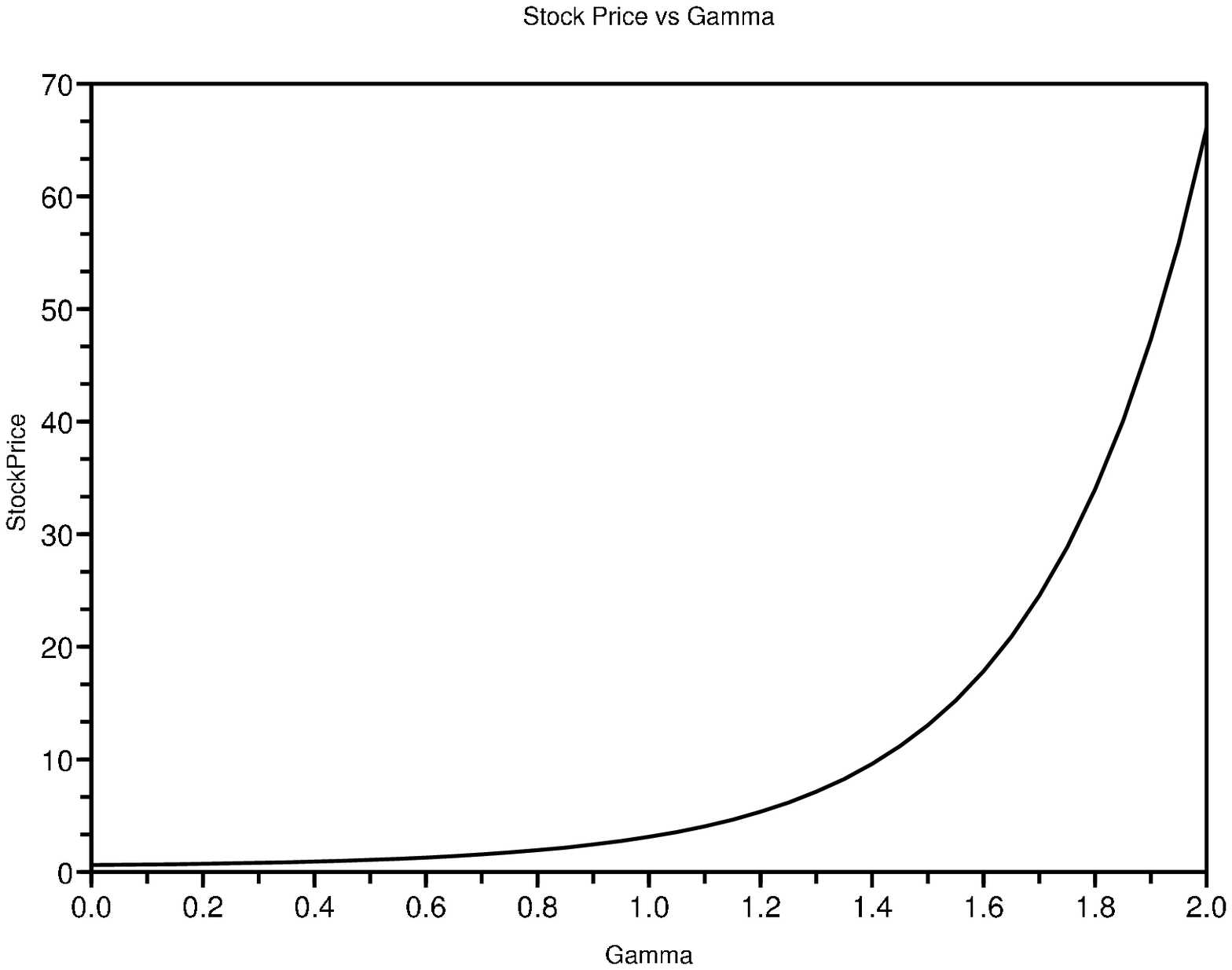}
  \end{center}
  \label{DBOUGraphStvsgamma}
\end{figure}

Figure \ref{DBOUGraphStvsgamma} may at first seem surprising, since it shows that the stock price is increasing in the risk aversion, $\Gamma$.  However, we recall that all agents have a CARA utility and furthermore, the parameters of our model are chosen so that the dividend process is non-negative. On the one hand, a larger value of $\Gamma$ means that the value of the dividend process becoming larger are valued more highly than before.  The downside of holding the stock is limited, since the dividend process is always non-negative.  This explains the behaviour shown in \ref{DBOUGraphStvsgamma}.

\begin{figure}[p]
  \caption{Volatility Surface}
  \begin{center}
    \includegraphics[width=1.0\textwidth]{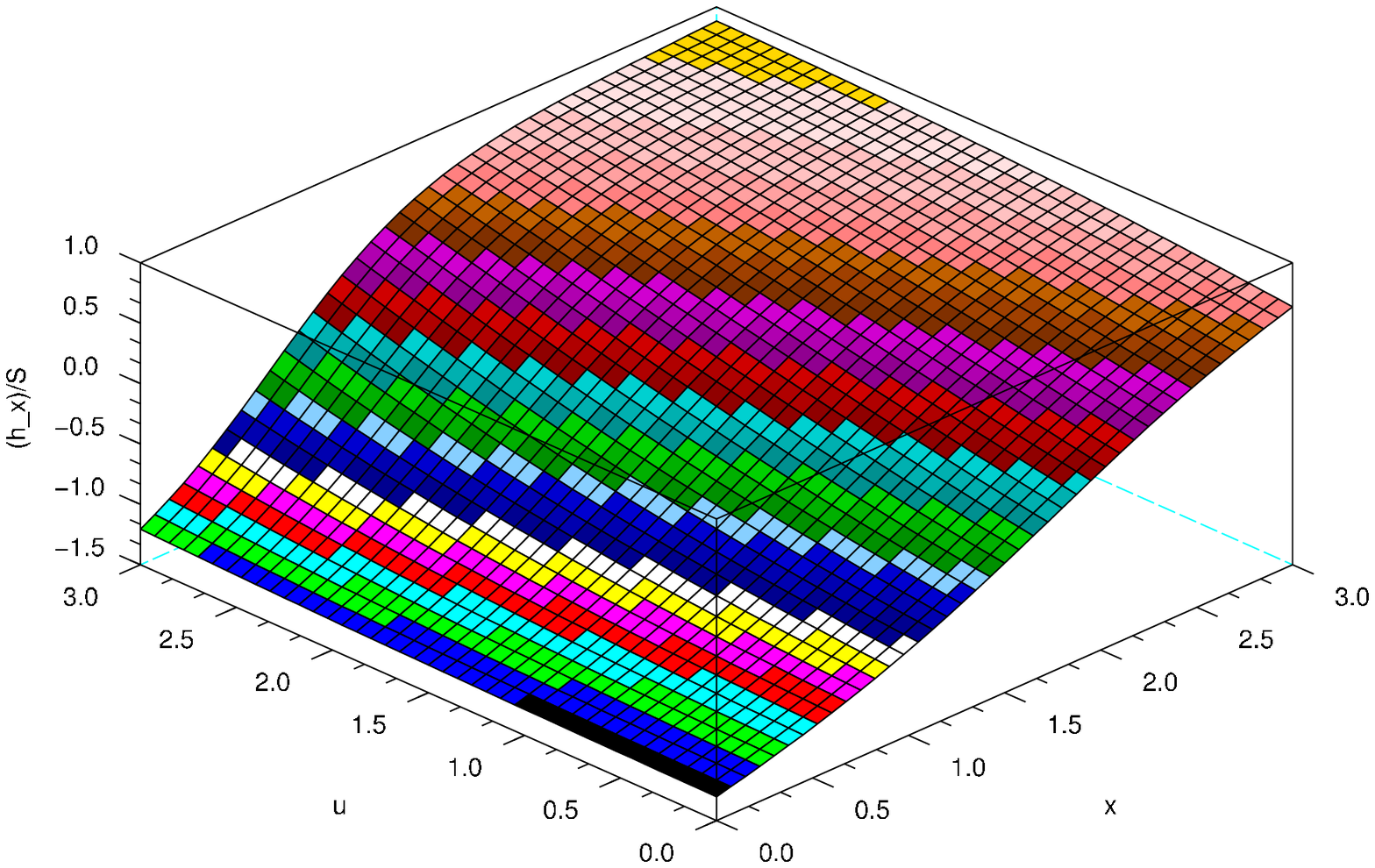}
  \end{center}
  \label{DBOUGraphVolsurf}
\end{figure}

The volatility surface\footnote{Note that the plot shows $h_{x}/S_{t}$; the absolute value of this would give the volatility.} in Figure \ref{DBOUGraphVolsurf} shows that the volatility appears to be increasing in both $X_{t}$ and $U_{t}$.  This seems reasonable:  if the dividend process has been varying greatly in the past, then $U_{t}$ will be large, and in this case we would expect the stock to have a larger volatility.  Also, if $X_{t}$ is small and $U_{t}$ is large, it means that either the dividend has been varying greatly, or the value of $X_{t}$ is abnormally small, so in these cases a large volatility should not be surprising.  However, increasing $X_{t}$ means that $U_{t}$ no longer implies that the dividend process has been varying so much - it just tells us that $X_{t}$ is typically large.  This explains why increasing $X_{t}$ will decrease the volatility.

\section{Conclusions}\label{DBOUCaseConclusions}

We have introduced a new model in which the dividend of the stock obeys an OU process for which none of the agents know the mean.  We derived a state price density and were able to use this to price the stock and a bond.  We also were able to deduce an interest rate model.  We produced graphs which illustrated the dependence of the stock price on the various parameters.   The behaviour shown in these graphs seemed very reasonable.  We also looked at how the parameter certainty case could be viewed as a special limit of the parameter uncertainty case. 

Extensions to this work include using a different utility function for the agents; a CRRA utility would be a natural choice.  In section \ref{DBOUContinuumAgents} we also had to assume a quite specific form for the distribution of the lifetimes of the agents. An obvious improvement would be to consider the problem with a different distribution of lifetimes, in particular one that did not depend on the parameters of the dividend process.  Unfortunately both these generalisations appear to make the calculations intractable.  
\bibliography{references}
\bibliographystyle{jmb}

\newpage

\appendix
\begin{center}
  {\bf APPENDIX}
\end{center}

\section{Stochastic Integrals}\label{DBOUCaseStochInts}

\subsection{Calculating $\xi_{t}$}

Recall that $\xi_{t}$ is given by:
\begin{align*}
\xi_{t} &=\int_{0}^{\infty} (W_{t} - W_{t-u}) \lambda e^{-\lambda u} du 
\end{align*}
By change of variables,
\begin{align*}
\xi_{t} & = W_{t} -  e^{-\lambda t} \int_{-\infty}^{t} \lambda e^{\lambda s} W_{s} ds
\end{align*}
So substituting from (\ref{DBOUWdefn}) gives: 
\begin{align}
\xi_{t} & = W_{t} + X_{0} -  e^{-\lambda t}\Big[ \int_{-\infty}^{t} X_{s} \lambda e^{\lambda s} ds +  \int_{-\infty}^{t}  \lambda e^{\lambda s} \int_{0}^{s} \lambda X_{r} dr  ds \Big]  \label{DBOUxitexpression2}
\end{align}
But the final term in the above expression is:
\begin{multline*}
- e^{-\lambda t}  \int_{-\infty}^{t}  \lambda e^{\lambda s} \int_{0}^{s} \lambda X_{r} dr  ds \\
=  e^{-\lambda t}  \int_{s=-\infty}^{0} \int_{r=s}^{0} \lambda e^{\lambda s} \lambda X_{r} dr ds - e^{-\lambda t} \int_{s=0}^{t} \int_{r=0}^{s} \lambda e^{\lambda s } \lambda X_{r} dr ds
\end{multline*}
Applying Fubini, we obtain:
\begin{align*}
e^{-\lambda t}  \int_{r=-\infty}^{0} \int_{s=-\infty}^{r} \lambda e^{\lambda s} \lambda X_{r} ds dr - e^{-\lambda t} \int_{r=0}^{t} \int_{s=r}^{t} \lambda e^{\lambda s } \lambda X_{r} ds dr
\end{align*}
Computing the integral with respect to $s$ gives:
\begin{align*}
& e^{-\lambda t} \big[\int_{-\infty}^{0} \lambda e^{\lambda r} X_{r} dr - e^{\lambda t} \int_{0}^{t} X_{r} \lambda dr + \int_{0}^{t} \lambda e^{\lambda r} X_{r} dr  \big] \\
& = e^{\lambda t} \int_{-\infty}^{t} \lambda e^{\lambda r} X_{r} dr - \int_{0}^{t} \lambda X_{r} dr
\end{align*}
Substituting this into (\ref{DBOUxitexpression2}) gives:
\begin{align*}
\xi_{t}=W_{t} +X_{0} - \int_{0}^{t} \lambda X_{r} dr 
\end{align*}
But recalling (\ref{DBOUWdefn}), we obtain:
\begin{align*}
\boxed{\xi_{t} = X_{t}}
\end{align*}

\subsection{Calculating $\eta_{t}$}
Recall that $\eta_{t}$ is given by: 
\begin{align*}
\eta_{t} &=\int_{0}^{\infty} (W_{t}- W_{t-u})^{2} \lambda e^{-\lambda u} du
\end{align*}
Changing variables we obtain:
\begin{align*}
\eta_{t} = e^{-\lambda t} \int_{-\infty}^{t} (W_{t}-W_{r})^{2} \lambda e^{\lambda r} dr
\end{align*}
Substituting from (\ref{DBOUWdefn}) gives:
\begin{align*}
 \eta_{t}  = e^{-\lambda t} \int_{-\infty}^{t} \big[(X_{t}-X_{r}) +\int_{r}^{t}\lambda X_{s} ds  \big]^{2} \lambda e^{\lambda r} dr 
\end{align*}
\begin{multline}
 = e^{-\lambda t} \int_{-\infty}^{t} (X_{t}-X_{r})^{2} \lambda e^{\lambda r} dr +  2 e^{-\lambda t} \int_{-\infty}^{t} (X_{t}-X_{r}) \big(\int_{r}^{t}\lambda X_{s} ds  \big) \lambda e^{\lambda r} dr \\
+ e^{-\lambda t} \int_{-\infty}^{t} \big(\int_{r}^{t} \lambda X_{s} ds  \big)^{2} \lambda e^{\lambda r} dr \label{DBOUetatexpression2}
\end{multline} 
We will now apply Fubini to two of these terms to deduce an expression for $\eta_{t}$.  Firstly, we work on:
\begin{align*}
\int_{r=-\infty}^{t} X_{t} \int_{s=r}^{t} \lambda X_{s} ds \lambda e^{\lambda r} dr
\end{align*}
By applying Fubini, we obtain:
\begin{align*}
& \int_{s=-\infty}^{t} X_{t} X_{s} \int_{r=-\infty}^{s} \lambda^{2} e^{\lambda r} dr ds \\
& = \int_{-\infty}^{t} X_{t} X_{s}  \lambda e^{\lambda s} ds
\end{align*}
Putting this into (\ref{DBOUetatexpression2}) gives: 
\begin{multline}
\eta_{t} = X_{t}^{2} +  e^{-\lambda t} \int_{-\infty}^{t} \lambda e^{\lambda r} X_{r}^{2} dr \\
- 2e^{-\lambda t} \int_{-\infty}^{t} X_{r} \big(\int_{r}^{t} \lambda X_{s} ds  \big) \lambda e^{\lambda r} dr \\
+ e^{-\lambda t} \int_{-\infty}^{t} \int_{r}^{t} \lambda X_{s} ds \int_{r}^{t} \lambda X_{v} dv \lambda e^{\lambda r} dr \label{DBOUetatexpression3}
\end{multline} 
The final term is:
\begin{align*}
2 e^{-\lambda t} \int_{r=-\infty}^{t} \int_{s=r}^{t} \int_{v=s}^{t} \lambda X_{s} \lambda X_{v} \lambda e^{\lambda r} dv ds dr
\end{align*}
where we have halved the area of integration in the $dv ds$ integral.  Applying Fubini yields:
\begin{align*}
& 2 e^{-\lambda t} \int_{s=-\infty}^{t} \int_{v=s}^{t} \int_{r=-\infty}^{s} \lambda X_{s} \lambda X_{v} \lambda e^{\lambda r} dr dv ds \\
&= 2 e^{-\lambda t} \int_{s=-\infty}^{t} \int_{v=s}^{t} \lambda X_{s} \lambda X_{v} e^{\lambda s} dv ds \\
&= 2 e^{-\lambda t} \int_{r=-\infty}^{t} \lambda X_{r} e^{\lambda r} \int_{s=r}^{t} \lambda X_{s} ds dr
\end{align*}
Substituting this into (\ref{DBOUetatexpression3}) gives:
\begin{align*}
\boxed{\eta_{t} = X_{t}^{2} +  e^{-\lambda t} \int_{-\infty}^{t} \lambda e^{\lambda s} X_{s}^{2} ds}
\end{align*}

\end{document}